\documentclass[twocolumn]{article}

\usepackage{geometry}
\usepackage[dvipsnames]{xcolor}
\usepackage{soul}
\usepackage{cuted}

\usepackage{subcaption}
\usepackage{booktabs}
\usepackage{adjustbox}
\usepackage{multirow}
\usepackage[all]{nowidow}
\usepackage{amsmath}
\usepackage{amssymb}
\usepackage{bm}
\usepackage{nicefrac}
\usepackage{mathtools}

\usepackage[separate-uncertainty=true,parse-numbers=true]{siunitx}

\usepackage{pgfplots}
\usepackage{tikz}
\pgfplotsset{compat = newest}
\usetikzlibrary{quotes,angles,calc,arrows,patterns,quotes,external}
\usepgfplotslibrary{groupplots}

\usepackage[colorlinks=true, citecolor=blue,urlcolor=blue,linkcolor=blue,filecolor=black]{hyperref}
\usepackage[capitalise,nameinlink,noabbrev]{cleveref} 

\usepackage{authblk}

\usepackage{algorithm}
\usepackage{algpseudocode}

\usepackage{multibib}
\newcites{Supplement}{Supplementary Information References}

\usepackage[T1]{fontenc}
\usepackage[utf8]{inputenc}

\usepackage{nowidow}
\usepackage{braket}

\title{Gigahertz-clocked Generation of Highly Indistinguishable Photons at C-band Wavelengths}

\author[1]{Robert Behrends}
\author[1,$\dagger$]{Lucas Rickert}
\author[1]{Nils D. Kewitz}
\author[1]{Martin v. Helversen}
\author[1]{Pratim K. Saha}
\author[1]{Mareike Lach}
\author[2]{Jochen Kaupp}
\author[2]{Yorick Reum}
\author[3]{Tobias Huber-Loyola}
\author[2]{Sven Höfling}
\author[2]{Andreas Pfenning}
\author[4,*]{Tobias Heindel}
\affil[1]{Institute of Physics and Astronomy, Technical Univeristy Berlin, Hardenbergstraße 36, 10623 Berlin, Germany
}
\affil[2]{University of Würzburg, Physikalisches Institut and Würzburg-Dresden Cluster of Excellence ctd.qmat, Lehrstuhl für Technische Physik, Germany}
\affil[3]{Institute of Photonics and Quantum Electronics (IPQ) and Center for Integrated Quantum Science and Technology (IQST), Karlsruhe Institute of Technology, Engesserstr. 5, 76131 Karlsruhe, Germany}
\affil[4]{Department for Quantum Technology, Univeristät Münster, Heisenbergstraße 11, 48149 Münster, Germany}
\affil[*]{Corresponding author: tobias.heindel@uni-muenster.de}
\date{\today}

\begin{document}


\twocolumn[
\begin{@twocolumnfalse}
\maketitle
\begin{abstract}
\noindent
High-performance single-photon sources at telecom C-band wavelentghs are key building blocks for applications in long-distance quantum communication. Here, we report the generation of highly indistinguishable, single photons at a clock-rate of 2.5\,GHz. This is achieved by coherently driving the biexciton transition ($T_1^\mathrm{XX}=64(1)\,$ps) of a semiconductor quantum dot embedded in a microcavity with strong asymmetric Purcell enhancement. Employing pulsed two-photon resonant excitation, strong multiphoton suppression with $g^{(2)}(0) < 4\%$ and high two-photon-interference visibility of $V_\mathrm{raw}> 85\%$ is observed. The observed photon indistinguishability is close to the theoretical limit expected for the photonically engineered radiative cascade and matches values obtained at lower repetition rates. Our results show a substantial advancement towards interference-based quantum information protocols at unprecedented data rates in the telecom C-Band.
\end{abstract}
\hspace{2cm}
\end{@twocolumnfalse}
]

\section{Introduction}
\label{sec:Intro}

Semiconductor quantum dots (QDs)~\cite{vajner_quantum_2022,heindel_quantum_2023} are at the forefront of solid-state based platforms for state-of-the-art quantum light sources in terms of multiphoton suppression~\cite{schweickert_-demand_2018}, two photon indistinguishability~\cite{zhai_quantum_2022} and end-to-end efficiency~\cite{tomm_bright_2021,ding_high-efficiency_2025}. While all of these benchmarks were achieved at near infrared wavelengths below \SI{1}{\micro\metre}, improvements in QD growth over the past decade yielded single photon emission at telecom C-band wavelengths~\cite{miyazawa_single-photon_2005,benyoucef_telecom-wavelength_2013,muller_quantum_2018,paul_single-photon_2017,phillips_purcell-enhanced_2024}. This is highly desirable for long haul applications requiring lowest attenuation in existing deployed fiber backbones \cite{Holewa_review_2025}.

A particular challenge for C-band emitting QDs was the generation of indistinguishable photons for InAs/GaAs QDs with metamorphic buffers~\cite{nawrath_resonance_2021,joos_coherently_2024}, and InAs/InP QDs~\cite{vajner_-demand_2024}, which has recently been overcome by optimized growth in the InAs/InGaAlAs/InP material system in combination with Purcell-enhancement~\cite{kim_two-photon_2025, michl2025spinphotoninterfacetelecomcband}. Using these devices, two-photon interference (TPI) visibilities of $\sim$90\% have been reported very recently employing coherent LA-phonon assisted excitation~\cite{hauser_2025} as well as two-photon resonant excitation (TPE) \cite{behrends_indistinguishable_2026} have been achieved at standard excitation repetition rates of 80\,MHz and 100\,MHz, respectively. At shorter wavelengths, on the other hand, first quantum optical studies have been conducted at clock-rates $\geq1$\,GHz \cite{schlehahn_electrically_2016, Yang2024}, using intrinsically fast decaying GaAs/AlGaAs QDs at $\sim780$\,nm~\cite{hopfmann_maximally_2021} and strongly Purcell-enhanced InAs/GaAs QDs at $\sim920$\,nm~\cite{rickert_high_2025,rickert_fiber-pigtailed_2025}, both reporting high photon indistinguishabilities in that spectral range. Such high-clock-rate experiments, which have been illusive in the telecom wavelength range until now, become crucial for advancing the field towards applications in high-performance quantum networks.

In this work, we report the GHz-clocked generation of single photons  with high photon indistinguishability in the telecom C-band using a Purcell-enhanced QD with strong asymmetric Purcell enhancement. Using TPE at a clock rate of 2.5\,GHz, we observe multiphoton suppression $<4\%$, with raw TPI visibilities $>85\%$, paving the way for fiber-compatible interference-based quantum information protocols at unprecedented rates.

\begin{figure*}[ht]
  \centering
  \includegraphics[width=1\textwidth]{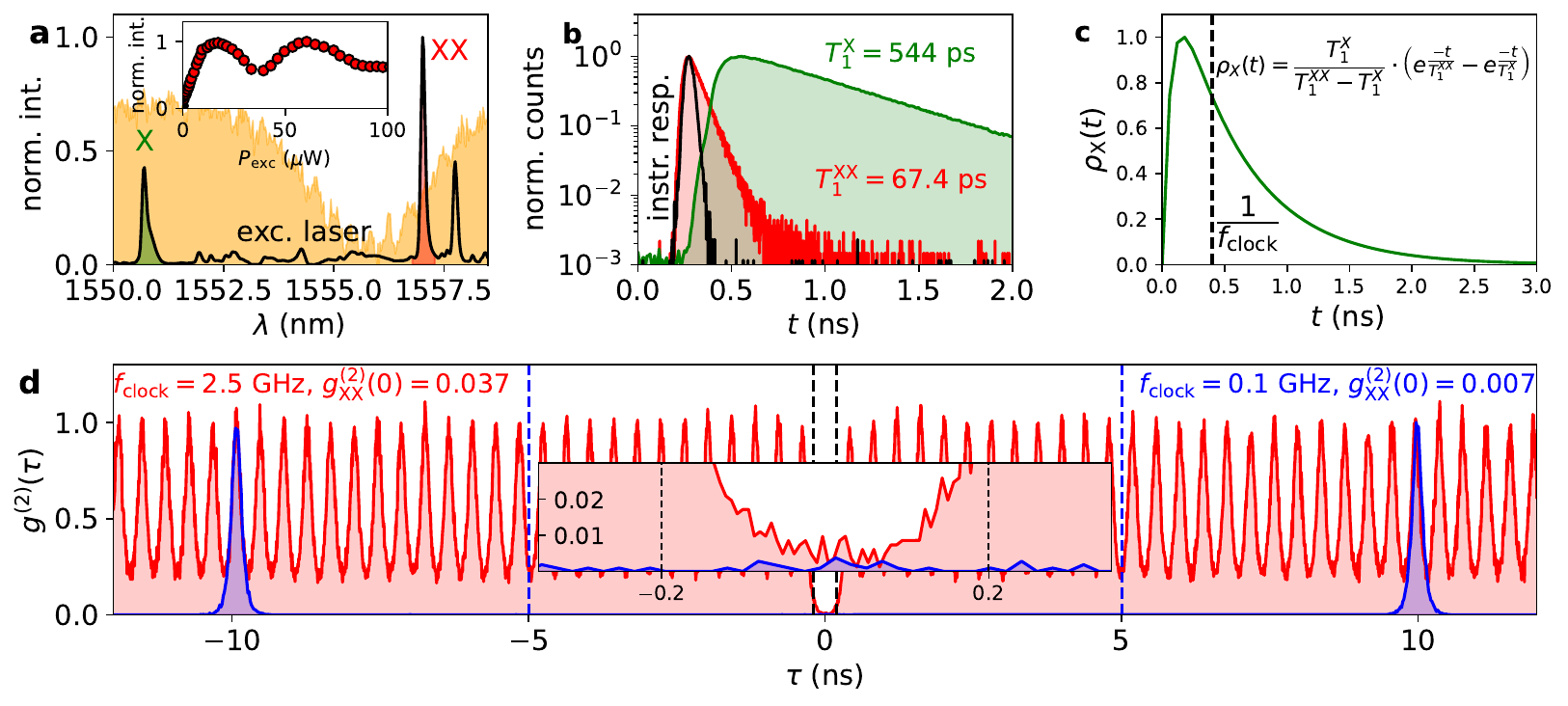}
   \caption{(a) QD spectrum under 2.5\,GHz clocked TPE. Neutral exciton (X), neutral biexciton (XX), as well as the excitation laser wavelength are indicated. The cavity mode is depicted in orange. The inset shows the power dependent photoluminescence (PL) signal of the XX state under 2.5\,GHz clocked TPE revealing Rabi rotations. (b) Time resolved X and XX PL signal with decay times extracted from mono-exponential fits. (c) Theoretical time-dependent X emission $\rho_\mathrm{X}(t)$ based on convoluted XX and X decay. The dashed line indicates the time interval between the excitation pulses. (d) Second order autocorrelation function $g^{(2)}(\tau)$ of the XX state under 2.5\,GHz (red) and 0.1\,GHz (blue)  clocked TPE, with integrated respective $g^{(2)}(0)$-values. The inset shows a zoom-in of the histogram around $\tau=0$ with indicated integration area for the GHz-clocked measurement indicated as black dashed lines. }
  \label{fig:Fig1}
\end{figure*}

\section{Results}
\label{sec:results}

The InAs/$\mathrm{In_{0.53}Al_{0.23}Ga_{0.24}As}$/InP QD used in this work is grown by molecular-beam epitaxy and integrated in a hybrid circular Bragg grating resonator. More information on sample growth and device fabrication can be found in Refs. \cite{kim_two-photon_2025, kaupp_purcellenhanced_2023}. To enable GHz clock-speed in our experiments, two ingredients are essential: firstly, a QD excitonic state with sufficiently fast emission decay time and, secondly, a high repetition rate laser is required. For pumping the QD at GHz rates, we used a pulsed laser with a repetition rate of $f_\mathrm{rep}=2.5$\,GHz, a  central emission wavelength of $\sim1560\,$nm, and $<250\,$fs pulse width. TPE is implemented using a pulse slicer to shape pulses with half of the $\ket{\mathrm{G}}$-$\ket{\mathrm{XX}}$ transition energy, where $\ket{\mathrm{G}}$ is defined as the ground state.
Additional reference experiments at a lower repetition rate were performed using a fiber-laser with $f_\mathrm{rep}=0.1$\,GHz. 
Figure~\ref{fig:Fig1}(a) shows the photoluminescence (PL) spectrum of the QD device under 2.5\,GHz clocked TPE, with the XX emission line at $\sim1557$\,nm. 
\begin{figure*}[!t]
  \centering
  \includegraphics[width=0.98\textwidth]{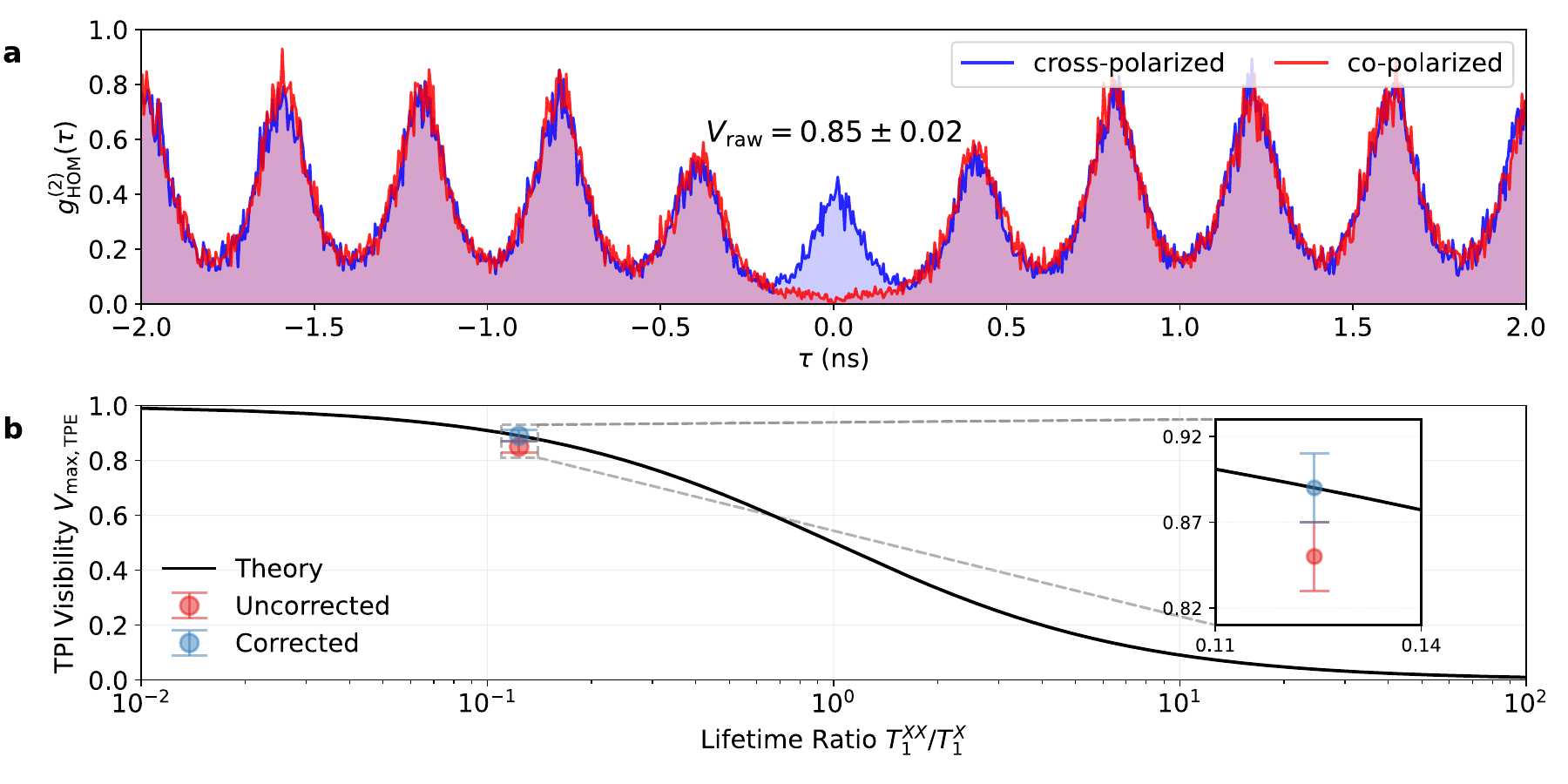}
   \caption{(a) Hong-Ou Mandel (HOM) autocorrelation measurement $g^{(2)}_\mathrm{HOM}(\tau)$ of the XX state under 2.5\,GHz clocked TPE. The photon time delay corresponds to $1/f_\mathrm{rep}=400\,$ps. The maximally distinguishable cross-polarized measurement is shown in blue, the maximally indistinguishable co-polarized measurement in red. (b) Dependence of TPI visibility on the radiative lifetime ratio. The solid black curve represents the theoretical maximum visibility $V_{\mathrm{max, TPE}}$ as a function of the ratio between biexciton ($T_1^{\rm XX}$) and exciton ($T_1^{\rm X}$) lifetimes. The experimental corrected result (blue circle) shows excellent agreement with the theoretical limit, while the uncorrected result (red circle) shows a slight deviation from the theoretical maxiumum, due to the residual pulse overlap.} 
  \label{fig:Fig2}
\end{figure*}
The inset shows the PL signal of the XX state as a function of excitation power $P_\mathrm{exc}$, where Rabi oscillations up to a pulse area of 3$\pi$ prove the coherent nature of the excitation.\\
For all subsequent measurements, the detected photons are spectrally filtered using three consecutive notch filters ($\lambda_\mathrm{c}=1555$\,nm, $\Delta\lambda=0.1$\,nm, OD3) based on a volume Bragg grating, followed by a spectrometer resulting in a spectral detection window of 70\,$\mu$eV (0.14 nm). Fitting the emission line yields a resolution-limited full width half maximum (FWHM) of 36 µeV (0.07 nm). This setup filters the phonon sideband of the XX transition but leaves its zero phonon line unfiltered.
The decay dynamics of X and XX are shown as time-resolved PL measurements in Fig.~\ref{fig:Fig1}(b), which yields decay times of $\approx 544$\,ps and $\approx 68$\,ps for the X- and XX-state, respectively. Note, that the latter is still well resolved by our detection system, with an instrument response function of 57.9\,ps FWHM. Deconvoluting with the detector resolution yields a radiative lifetime of the XX-state of $T_{1}^{\rm XX}=64(1)$\,ps. We achieve a detected photon rate of 2.2\,MHz, representing a $\approx$12-fold increase in the count rate under 2.5\,GHz clocked excitation, compared to 100\,MHz clocked excitation. Fig.~\ref{fig:Fig1}(c) shows the theoretical time-resolved X emission as a convolution of both XX and X exponential decays~\cite{sbresny_stimulated_2022} as
\begin{equation}
    \rho_\mathrm{X}(t) = \dfrac{T_1^\mathrm{X}}{T_1^\mathrm{XX}-T_1^\mathrm{X}}\cdot \left(e^\frac{-t}{T_1^\mathrm{XX}} - e^\frac{-t}{T_1^\mathrm{X}}\right).
\end{equation}
By integrating these emission events within the $1/f_\mathrm{clock}=400\,$ps time-window and comparing to the remaining events on infinite time-scales, we can determine the fraction of still populated $\ket{\mathrm{X}}$-states, which prevent the population of $\ket{\mathrm{XX}}$-states for the following excitation pulses at 2.5\,GHz clock-rate. 
Further, we obtain a theoretical possible increase in count-rate between 100\,MHz and 2.5\,GHz excitation rate by a factor of 13.5,  assuming a non-ideal preparation fidelity of 80\%, which is in good agreement with the above mentioned 12-fold observation in the experiment. We thus attribute the limited gain mostly to the slow radiative decay of the intermediate X state, which prevents the excitation of the cascade if the system has not fully relaxed to the ground state between pulses. The remaining discrepancy could stem from altered blinking-contributions or a further reduced preparation fidelity at these high clock-rates.

Fig.~\ref{fig:Fig1}(d) shows the second order autocorrelation $g^{(2)}(\tau)$ obtained for the XX photons in Hanbury Brown-Twiss (HBT) experiments under TPE at 2.5\,GHz (red lines) and 0.1\,GHz TPE (blue lines), respctively. The antibunching values $g^{(2)}(0)$ are extracted from the coincidence histograms by comparing the respective areas $A_0$ in the time interval $\pm200$\,ps  to the average peak height of the surrounding peaks $A_n, n\neq0$. 
From the blinking corrected data (see Supplementary Information (S.I.)), an integrated multiphoton suppression of $g_\mathrm{XX}^{(2)}(0)=(3.7\pm0.7)\%$ is obtained for the $f_\mathrm{rep}=2.5$\,GHz, while the slower $f_\mathrm{rep}=0.1$\,GHz measurements yields a value of $g_\mathrm{XX}^{(2)}(0)=(0.7\pm0.1)\%$. 
The higher multiphoton contribution at GHz clock-rates is mainly attributed to residual pulse overlap in the time-bin $\tau=0^\mathrm{+0.2\,ns}_\mathrm{-0.2\,ns}$, due to the finite XX decay time. 
Further, we want to point out that by analyzing the blinking behaviour, we extract an off-on ratio of 16.6(7)\% between the two states of the investigated QD causing the blinking effect, which further reduces the overall extraction efficiency.
The second-order autocorrelation measurement confirms, that the cavity enhanced XX transition maintains its excellent single photon emitter properties also under GHz optical pump. Moreover, the values are comparable to integrated $g^{(2)}(0)$-values reported for quasi-resonantly excited InAs/GaAs QDs~\cite{rickert_high_2025,rickert_fiber-pigtailed_2025} at slightly lower optical repetition rates of $f_\mathrm{rep}=1.28\,$GHz. Interestingly, since there are no other comparable reports on optical excitation at GHz rates in the telecom C-band to our knowledge, our observed $g^{(2)}(0)$-values clearly surpass the performance of QD devices using electrical GHz excitation, both at short~\cite{hargart_electrically_2013,schlehahn_electrically_2016} and C-band wavelengths~\cite{anderson_gigahertz-clocked_2020,shooter_1ghz_2020}, highlighting the advantages of coherent optical excitation compared to off-resonant electrical pumping. Also note, that previous work on GHz-clocked TPE using non-Purcell-enhanced GaAs QDs reported multiphoton suppression values of 0.3\%~\cite{hopfmann_maximally_2021}, which was obtained at less than half the excitation frequency of our experiments.
Remaining limitations in the current multiphoton suppression for the QDs investigated here might arise from the remaining dark counts on the used telecom SNSPDs, as well as uncorrelated photons from the cavity mode in close resonance to the XX transition. We further note that pioneering early work on GHz-clocked QD single-photon sources even reported $g^{(2)}(0)=0$~\cite{stock_high-speed_2011}, the simulations underlying this claim, however, lacked a consistent validation from our viewpoint. \\
In order to check for the indistinguishability of the emitted photons under 2.5\,GHz TPE, the two-photon interference (TPI) visibility $V$ of consecutively emitted photons is probed in Hong-Ou-Mandel-type experiments~\cite{hong_measurement_1987}, where the time delay between the two photons is set to $1/f_\mathrm{rep}=0.4$\,ns. Figure~\ref{fig:Fig2}(a) presents the resulting autocorrelation histograms $g_\mathrm{HOM}^{(2)}(\tau)$ in co- and cross-polarized measurement configuration. Both histograms are corrected for blinking to ensure proper normalization (see S.I.).
The raw, blinking corrected TPI visibility from the integrated events in the time-interval $\pm1/f_\mathrm{rep}$, of co- and cross-polarized measurements is obtained according to Eq.~\ref{eq:V_raw}
\begin{equation}
    V_\mathrm{raw} = 1 - \frac{A_\mathrm{co}(\tau=0)}{A_\mathrm{cross}(\tau=0)} = 0.85(2)
    \label{eq:V_raw}
\end{equation}
%
As discussed above in the context of $g^{(2)}(\tau)$, this raw value is impaired by the finite overlap of neighboring coincidence pulses within $\pm1/f_\mathrm{rep}$ due to the high clock-speed. Accounting for this in the HOM measurement analysis, we fit the data with a sum of exponential functions, modeling both the central peak and the adjacent side peaks, to quantify their contribution to their respective neighboring time bins (see S.I.). We obtain an extracted corrected visibility of $\mathrm{V_{corr} = 0.89(2)}$. As shown in Fig.~\ref{fig:Fig2} (b), this value is in excellent agreement with the theoretical limit of $V_\mathrm{theo}\approx0.89$ expected for the decay time ratio of XX and X state extracted from our experimental data in Fig~\ref{fig:Fig1}(b)~\cite{simon_creating_2005,scholl_crux_2020}.
The TPI results reported above surpass previous measurements on QD indistinguishability both in employed excitation clock-rate and observed visibility~\cite{hopfmann_maximally_2021,rickert_high_2025,rickert_fiber-pigtailed_2025}, while also achieving these performances for fiber-compatible C-band emission.

\section{Conclusions}
\label{sec:conclusions}

In summary, we demonstrated the generation of highly indistinguishable single-photons at 2.5\, GHz optical clock-rate from a semiconductor QD directly emitting in the telecom C-band. Employing TPE at the $\pi$-pulse, we achieved a multiphoton suppression of $g_\mathrm{XX}^{(2)}(0)=(3.7\pm0.7)\%$, a raw TPI visibility of $V_\mathrm{raw}>85\%$, and a corrected TPI visibility matching the theoretical limit. This confirms that high photon indistinguishability from QD sources are maintained under unprecedented clock-speeds, showing up prospects for substantial leaps in the performance of QD-based telecom quantum light sources for interference-based photonic quantum information applications. 

Remaining limitations in our study concern the residual pulse overlap due to the finite XX lifetime, affecting the multi-photon suppression as well as the TPI visibility, and the relatively slow X decay, limiting the achievable single photon flux. To overcome these limitations in future work, efforts may focus on the following two routes: Route~A could focus on achieving still an asymmetric Purcell-enhancement of XX and X state with a factor of 10 or higher, e.g. by enhancing the XX state with a Purcell of 50, and the X state with a Purcell of 5. This would reduce the pulse overlap for the XX photons at the employed clock-rates, further increase the achievable count-rate by limiting the influence of the X intermediate state, and maintain the high TPI visibility under TPE excitation. These Purcell enhancements would require a determinstic integration technique with very high accuracy~\cite{rickert_high_2025,buchinger_deterministic_2025}. Route~B would focus on collecting the photons from a deterministically Purcell-enhanced X state with a Purcell factor >25 in combination with stimulated TPE~\cite{yan_double-pulse_2022, sbresny_stimulated_2022, wei_tailoring_2022, behrends_indistinguishable_2026} at $>$GHz optical excitation rates. This would negate the TPI limitations of the XX-X cascade independent of the XX state decay time and provide highly indistinguishable X-photons at GHz clock-rates with the benefit of a two-photon excitation process for reduced re-excitation and high multiphoton suppression.

\section{Acknowledgments}
We gratefully acknowledge expert sample fabrication by M. Emmerling and thank Menhir Photonics AG for providing the laser system used for the experiments in this work.

\section{Data and Availability}
The data presented in this work will be made available on a public repository upon peer-reviewed publication.

\section{Funding}
The authors acknowledge financial support by the German Federal Ministry of Research, Technology and Space (BMFTR) via the project “QuSecure” (Grant No. 13N14876) within the funding program Photonic Research Germany, the BMFTR joint projects “tubLAN Q.0” (Grant No. 16KISQ087K), the project QuNET+ICLink (Grant No. 16KIS1967) in the context of the federal government’s research framework in IT-security “Digital. Secure. Sovereign.” and the project "PhotonQ" (Grant No. 13N15759). The authors further acknowledge funding by the State of Bavaria. T. H.-L. acknowledges financial support by the BMFTR via the project "Qecs" (Grant No. 13N16272). A. T. P acknowledges funding by the BMFTR via the project "Ferro35" (Grant No. 13N17641).

\section{Disclosure}
The authors declare no conflict of interest.
\\

\noindent $^{\dagger}$ Current address: Toshiba Europe Ltd., 208~Cambridge Science Park, Milton Road, Cambridge, CB4~0GZ, United Kingdom

\bibliographystyle{ieeetr}
\bibliography{bibliography}

\end{document}